\documentclass[preprint,eqsecnum,aps,nofootinbib]{revtex4}
\usepackage{amsfonts,amsmath,amssymb,amsthm}
\usepackage{latexsym}
\usepackage{bbm,bm}
\usepackage{graphicx}

\newcommand{\ket}[1]{\lvert #1 \rangle}
\newcommand{\bra}[1]{\langle #1 \lvert}
\newcommand{\beq}{\begin{equation}}
\newcommand{\eeq}{\end{equation}}
\newcommand{\beqs}{\begin{eqnarray}}
\newcommand{\eeqs}{\end{eqnarray}}

\begin{document}

\title{Entanglement Degradation in the presence of $(4+n)$-dimensional Schwarzschild Black Hole}

\author{DaeKil Park$^{1,2}$}

\affiliation{$^1$ Department of Physics, Kyungnam University, Changwon, 631-701, Korea \\
             $^2$ Department of Electronic Engineering, Kyungnam University, Changwon, 631-701, Korea }

\begin{abstract}
In this paper we compute the various bipartite quantum correlations in the presence of the $(4+n)$-dimensional Schwarzschild 
black hole. In particular, we focus on the $n$-dependence of various bosonic bipartite entanglements. For the case between Alice and Rob, 
where the former is free falling observer and the latter is at the near-horizon region, the quantum correlation is degraded compared to 
the case in the absence of the black hole. The degradation rate increases with decreasing $n$. We also compute the physically
inaccessible correlations. It is found that there is no creation of quantum correlation between Alice and AntiRob. For the case 
between Rob and AntiRob the quantum entanglement is created although they are separated in the causally disconnected regions. 
It is found that contrary to the physically accessible correlation the entanglement between Rob and AntiRob decreases with increasing $n$. 
\end{abstract}

\maketitle

\section{Introduction}
It is evident that quantum information processes such as  quantum teleportation\cite{bennett93}, 
quantum cryptography\cite{cryptography}, and quantum computer\cite{qc} will play crucial role in the future technology.
In this reason, much attention is paid, recently, to the quantum entanglement\cite{horodecki07} because it is regarded 
as a genuine physical resource for the quantum information processing.

Although research into the quantum entanglement has a long history\cite{epr-35,schrodinger-35}, the study of its properties
in the relativistic setting was initiated recently\cite{peres1,peres2,peres3,inertial1,inertial2,inertial3,inertial4,noninertial1,noninertial2,noninertial3,
noninertial4,noninertial5,noninertial6,noninertial7,noninertial8,noninertial9,noninertial10,noninertial11,noninertial12}.
The main issue in this subject is to understand how a given entanglement is changed in the inertial-\cite{inertial1,inertial2,inertial3,inertial4}
and non-inertial-\cite{noninertial1,noninertial2,noninertial3,
noninertial4,noninertial5,noninertial6,noninertial7,noninertial8,noninertial9,noninertial10,noninertial11,noninertial12} frames.
In the non-inertial frame degradation of the entanglement occurs, which is related to the well-known Unruh effect\cite{unruh1,unruh2}.
More recently, moreover, the change of the entanglement in the black hole background was also examined\cite{black1,black2,black3}.

On the other hand, braneworld scenario\cite{brane1,brane2,brane3,brane4}, one of the modern cosmology, assumes that our $4d$ spacetime universe
is embedded in higher-dimensional world. In this reason, the absorption and emission properties of the higher-dimensional black holes were 
investigated few years ago\cite{higher1,higher2,higher3,higher4,higher5}.

The purpose of this paper is to examine the degradation of the bipartite entanglement in the presence of the $(4+n)$-dimensional 
Schwarzschild black hole. In order to explore the degradation we assume that Alice and Rob share the maximally entangled state initially.
After sharing Rob moves to the near-horizon region while Alice is free falling into the black hole. In this situation we will compute
the entanglement by adopting the negativity\cite{negativity} as an entanglement measure. Our main focus in this paper is to investigate the 
effect of the extra dimensions $n$ in the entanglement degradation. Of course, the result of Ref. \cite{black2} is reproduced when $n=0$.

The paper is organized as follows. In section II we review the spacetime geometry of the $(4+n)$-dimensional Schwarzschild black hole.
For later convenience we compare the $(4+n)$-dimensional Schwarzschild spacetime with the Rindler spacetime in this section. This comparison
enables us to transform our problem into the degradation of entanglement in the non-inertial frame, which was studied in Ref.  \cite{noninertial1,noninertial2,
noninertial3,noninertial4,noninertial5,noninertial6,noninertial7,noninertial8,noninertial9,noninertial10,noninertial11,noninertial12}. 
In section III we discuss on the $n$-dependence of the entanglement. The correlation between Alice and Rob is generally degraded in the presence
of the black hole. It is found that the degradation becomes weaker and weaker with increasing $n$. For the case between Rob and AntiRob the 
quantum correlation, contrary to the physically accessible correlation, decreases with increasing $n$.
In section IV a brief conclusion is given. In appendix A explicit computation of negativity is performed.

\section{Space-Time Geometry}

The higher-dimensional black hole solutions of the Einstein field equation
were discussed in detail in Ref.\cite{myers86}. The explicit expression of the
$(4+n)$-dimensional Schwarzschild black hole solution in terms of the usual Schwarzschild
coordinates is in the following:
\begin{equation}
\label{schwarz1}
ds^2 = -h(r) dt^2 + h^{-1}(r) dr^2 + r^2 d\Omega_{n+2}^2
\end{equation}
where
\begin{eqnarray}
\label{horizonp}
& &h(r) = 1 - \left(\frac{r_H}{r}\right)^{n+1}                        \\   \nonumber
& &d\Omega_{n+2}^2 = d\theta_{n+1}^2 +
\sin^2 \theta_{n+1} \Bigg( d\theta_{n}^2 + \sin^2 \theta_n \bigg(
\cdots + \sin^2 \theta_2 \left( d\theta_1^2 + \sin^2 \theta_1 d\phi^2
           \right) \cdots \bigg) \Bigg).
\end{eqnarray}
The horizon radius $r_H$ is related to the black hole mass $M$ as following:
\begin{equation}
\label{relation2}
r_H^{n+1} = \frac{8 \Gamma \left(\frac{n+3}{2}\right)}
                 {(n+2) \pi^{\frac{n+1}{2}}}
            \frac{M}{M_{\ast}^{n+2}}
\end{equation}
where $M_{\ast} = G^{-1/(n+2)}$ is a $(4+n)$-dimensional Planck mass and
$G$ is a Newton constant.

Because of the symmetry of the problem we restrict our attention to the temporal and radial coordinates. Thus, 
the part of the metric we will analyze is 
\begin{equation}
\label{metric2}
d \ell^2 = - h(r) dt^2 + h^{-1}(r) dr^2.
\end{equation}
We can write the line-element $d \ell^2$ in terms of the proper time $\tau$ of an observer located in $r = r_0$ as follows:
\begin{equation}
\label{metric3}
d \ell^2 = - \frac{h(r)}{h_0} d \tau^2 + h^{-1}(r) dr^2
\end{equation}
where
$h_0 \equiv h(r_0)$ and $d\tau = \sqrt{h_0} dt$. 

Now, we define a new spatial coordinate $z$ as follows:
\begin{equation}
\label{spatial1}
z^2 = \frac{4 r_H^{1 - n}}{(1 + n)^2} \bigg( r^{1+n} - r_H^{1+n} \bigg).
\end{equation}
Then, the profile function $h(r)$ can be written as 
\begin{equation}
\label{profile1}
h(r) = \frac{(\kappa z)^2}{1 + (\kappa z)^2}
\end{equation}
where $\kappa$ is a surface gravity defined by 
\begin{equation}
\label{surface-gravity}
\kappa \equiv \frac{1}{2} h' (r_H) = \frac{1+n}{2 r_H}.
\end{equation}
In terms of $z$ it is easy to show that the line-element $d\ell^2$ becomes
\begin{equation}
\label{metric4}
d\ell^2 = - \frac{1}{h_0} \frac{(\kappa z)^2}{1 + (\kappa z)^2} d\tau^2 + 
          \left[1 + (\kappa z)^2 \right]^{(1-n) / (1+n)} dz^2.
\end{equation}
In the near-horizon region, {\it i.e.} $r \sim r_H$, Eq. (\ref{metric4}) reduces to 
\begin{equation}
\label{metric5}
d\ell^2 \approx - \left( \frac{\kappa z}{\sqrt{h_0}} \right)^2 d\tau^2 + dz^2.
\end{equation}
This is just the metric of the Rindler space with the acceleration parameter $\kappa / \sqrt{h_0}$.

In order to discuss on the physical states we define the Kruskal coordinates
\begin{equation}
\label{kruskal}
\bar{u} = - \kappa^{-1} e^{-\kappa (t - r_*)}           \hspace{1.0cm}
\bar{v} = \kappa^{-1} e^{\kappa (t + r_*)}
\end{equation}
where the ``tortoise'' coordinate $r_*$ is defined by
\begin{equation}
\label{tortoise}
\frac{d r_*}{d r} = h^{-1} (r).
\end{equation} 
In terms of the Kruskal coordinates the line-element $d\ell^2$ becomes
\begin{equation}
\label{metric6}
d\ell^2 = -h(r) e^{-2\kappa r_*} d\bar{u} d\bar{v}.
\end{equation}
Since $r_*$ reduces to 
$$ r_* \sim r_H + \frac{r_H}{n+1} \ln \bigg| \frac{r}{r_H} - 1 \bigg|$$
at the near-horizon region, we can re-express $d\ell^2$ in this region as
$$ d\ell^2 \sim -(n+1) e^{-(n+1)} d\bar{u} d\bar{v}    \hspace{1.0cm}     \bar{u} \bar{v} = -\frac{z^2}{n+1}.$$
As authors in Ref. \cite{black2} suggested, there are three regions in this background, in which we can define the different time-like 
Killing vectors. Thus, one should define the physical vacuum in each region. First vacuum is known as the Hartle-Hawking vacuum
$\ket{0}_H$. This is defined by the time-like Killing vector which is proportional to $\partial_{\bar{u}} + \partial_{\bar{v}}$. 
Second and third vacua are the Boulware vacuum\cite{boulware75} $\ket{0}_B$ and anti-Boulware vacuum $\ket{0}_{\bar{B}}$, which are 
defined by the time-like Killing vectors $\partial_t$ and $-\partial_t$, respectively. 

Since the metric (\ref{metric5}) reduces to the Rindler space metric with the acceleration $\kappa / \sqrt{h_0}$ in the near-horizon region, 
one can derive the interrelations between these vacua using the Unruh effect\cite{unruh1,unruh2}. The relations beyond the single-mode 
approximation were derived in Ref. \cite{beyond}. Corresponding $\ket{0}_H$ to Minkowski vacuum and, $\ket{0}_B$ and $\ket{0}_{\bar{B}}$ to 
the vacua in the left- and right-wedges of Rindler space, one can derive the following relations:
\begin{eqnarray}
\label{inter1}
& &\ket{0_{\Omega}}_H = \frac{1}{\cosh r} \sum_{n=0}^{\infty} \tanh^n r \ket{n_{\Omega}}_B \ket{n_{\Omega}}_{\bar{B}}        \\    \nonumber
& &\ket{1_{\Omega}}_H = \frac{1}{\cosh^2 r} \sum_{n=0}^{\infty} \sqrt{n+1} \tanh^n r
\left[ q_L \ket{n_{\Omega}}_B \ket{(n+1)_{\Omega}}_{\bar{B}} + q_R \ket{(n+1)_{\Omega}}_B \ket{n_{\Omega}}_{\bar{B}} \right]
\end{eqnarray}
where $|q_L|^2 + |q_R|^2 = 1$ and 
\begin{equation}
\label{acceleration1}
\tanh r = \mbox{exp} \left( - \frac{\pi \sqrt{h_0} \Omega}{\kappa} \right).
\end{equation}
In Eq. (\ref{inter1}) $\ket{m_{\Omega}}_F$ means $m$-particle states of energy $\Omega$, which is constructed by operating the creation 
operator $m$ times to the vacuum $\ket{0}_F$ with $F = \{H, B, \bar{B} \}$. In next section we will use Eq. (\ref{inter1}) to analyze the 
bosonic entanglement degradation in the presence of the black hole (\ref{schwarz1}).

\section{Entanglement Degradation}

Let us consider a situation that Alice and Rob initially share the maximally entangled state\begin{equation}
\label{maximal}
\ket{\psi}_{AR} = \frac{1}{\sqrt{2}} \left( \ket{00}_H + \ket{11}_H \right)_{AR}
\end{equation}
when both are free falling into the black hole. Since $\partial_{\bar{u}} + \partial_{\bar{v}}$ is proportional to the time-like
Killing vector for the free falling observer, Eq. (\ref{maximal}) is understood to be written as a Hartle-Hawking basis. For simplicity, the energy
parameters $\Omega_A$ and $\Omega_R$ for Alice and Rob are ignored in Eq. (\ref{maximal}). 

Now, we assume that after sharing $\ket{\psi}_{AR}$, Rob moves to the near-horizon region, {\it i.e.} $r = r_0 \sim r_H$. Thus, 
we restrict ourselves into
\begin{equation}
\label{near1}
1 < R_S \equiv \frac{r_0}{r_H} < 1.05
\end{equation}
throughout this paper.
Therefore, Eq. (\ref{acceleration1}) implies
\begin{equation}
\label{statistical-1}
\tanh r = \exp \left( -\frac{\Omega}{2} \sqrt{1 - \frac{1}{R_s^{n+1}}} \right)
\end{equation}
where $\Omega \equiv 2\pi \Omega_R / \kappa$. In this assumption Eq. (\ref{inter1}) implies that $\ket{\psi}_{AR}$ is changed into
\begin{equation}
\label{black-1}
\ket{\psi}_{AR\bar{R}} = \frac{1}{\sqrt{2}} \sum_{n=0}^{\infty} \frac{\tanh^n r}{\cosh^2 r}
\bigg[ \cosh r \ket{0,n,n} + \sqrt{n+1} q_L \ket{1,n,n+1} + \sqrt{n+1} q_R \ket{1,n+1,n} \bigg].
\end{equation}

In this paper our main interest is to examine the effect of the extra dimension $n$ in the entanglement degradation. Thus, 
we will choose $q_R = 1$ for simplicity. Furthermore, as shown in Ref.\cite{noninertial1}, Eq. (\ref{inter1}) with $q_R = 1$ case corresponds 
to the Unruh transformation within the single-mode approximation. 

\subsection{Alice-Rob quantum correlations}
Now, let us discuss on the entanglement between Alice and Rob. 
In order to examine the entanglement in the presence of the $(4+n)$-dimensional black hole, we should compute the 
bipartite entanglement of a quantum state $\rho_{AR}$, where the subscript $AR$ stands for Alice and Rob. 
Although there are a lot of entanglement measures which quantify the bipartite entanglement 
such as entanglement of formation\cite{bennett96}, concurrence\cite{woot-98}, and relative entropy of entanglement\cite{vedral-97-1}, most of them 
are very hard to compute mainly due to that fact that $\rho_{AR}$ is a state in qudit system. Therefore, we choose a negativity\cite{negativity}
in this paper for tractable computation. The negativity ${\cal N}_{AR}$ of $\rho_{AR}$ is defined as 
\begin{equation}
\label{negativity1}
{\cal N}_{AR} = \frac{1}{2} \sum_i \left( |\lambda_i| - \lambda_i \right) = - \sum_{\lambda_i < 0} \lambda_i,
\end{equation}
where $\lambda_i$ are the eigenvalues of $\rho_{AR}^{T_A}$ and $T_A$ is a partial transposition with respect to the party $A$. 
It is worthwhile noting that the negativity ${\cal N}_{AR}$ in the absence of the black hole is $1/2$.

\begin{figure}[ht!]
\begin{center}
\includegraphics[height=6.5cm]{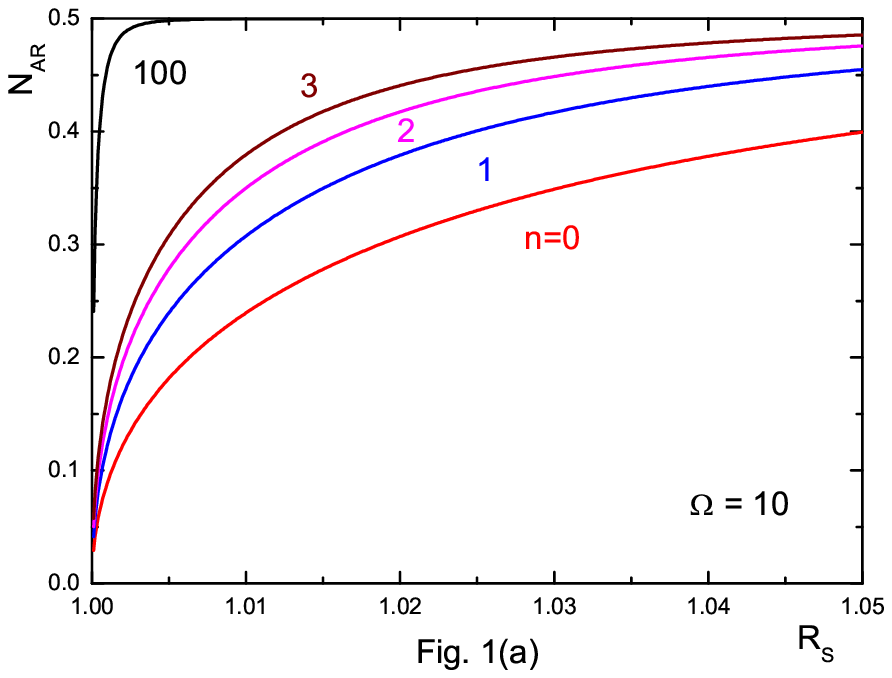}
\includegraphics[height=6.5cm]{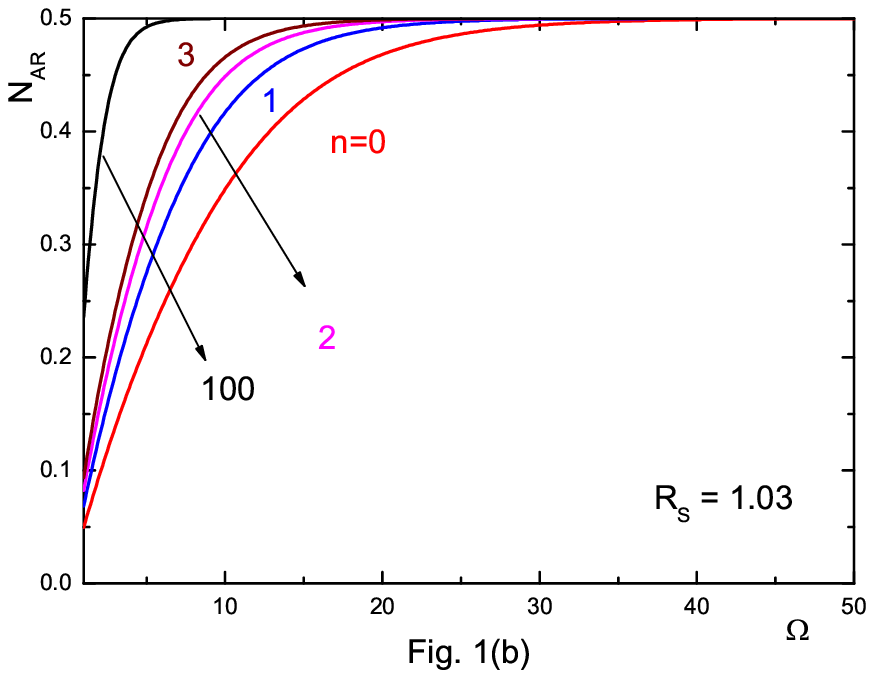}
\caption[fig1]{(Color online) In (a) we plot the $R_S$-dependence of 
${\cal N}_{AR}$ when $n=0, 1, 2, 3, 100$. We fix $\Omega$ as 
$\Omega = 10$. In (b) we plot the $\Omega$-dependence of ${\cal 
N}_{AR}$ when $n=0, 1, 2, 3, 100$. We fix $R_S$ as 
$R_S = 1.03$. 
}
\end{center}
\end{figure}

The computation of ${\cal N}_{AR}$ is described in Appendix A.
In Fig. 1 we plot ${\cal N}_{AR}$ with varying the number of extra dimensions as $n=0, 1, 2, 3, 100$. 
In Fig. 1(a) we plot the $R_S$-dependence of ${\cal N}_{AR}$ with choosing $\Omega = 10$. As this figure exhibits, ${\cal N}_{AR}$ is 
less than $0.5$, so that as expected, the degradation of the entanglement occurs. 
The rate of the degradation decreases with increasing $R_S$, which means that Rob goes away from the black hole. When Rob approaches
to the horizon, {\it i.e.} $r_0 \rightarrow r_H$, all bipartite entanglement vanishes. Another remarkable fact this figure shows is that the degradation becomes weaker with increasing $n$. This fact can be explained as follows. In $(4+n)$ dimension the Newtonian gravitational force is 
proportional to $1 / r^{n+2}$. Therefore, it goes weaker and weaker with increasing $n$. Thus, the effective acceleration Rob needs to remain 
outside the black hole becomes smaller with increasing $n$. It causes the weaker degradation of the entanglement between Alice and Rob.
In Fig. 1(b) we plot the $\Omega$-dependence of ${\cal N}_{AR}$ with choosing $R_S = 1.03$. This figure also shows that the degradation 
becomes weaker with increasing $n$. The reason for this fact can be explained same way.

\subsection{Alice-AntiRob quantum correlations}

The computation of negativity between Alice and AntiRob is given in appendix A. It is shown that no entanglement is created 
between these parties when $q_R = 1$, which corresponds to the Unruh transformation within the single-mode approximation.

\subsection{Rob-AntiRob quantum correlations}

\begin{figure}[ht!]
\begin{center}
\includegraphics[height=6.5cm]{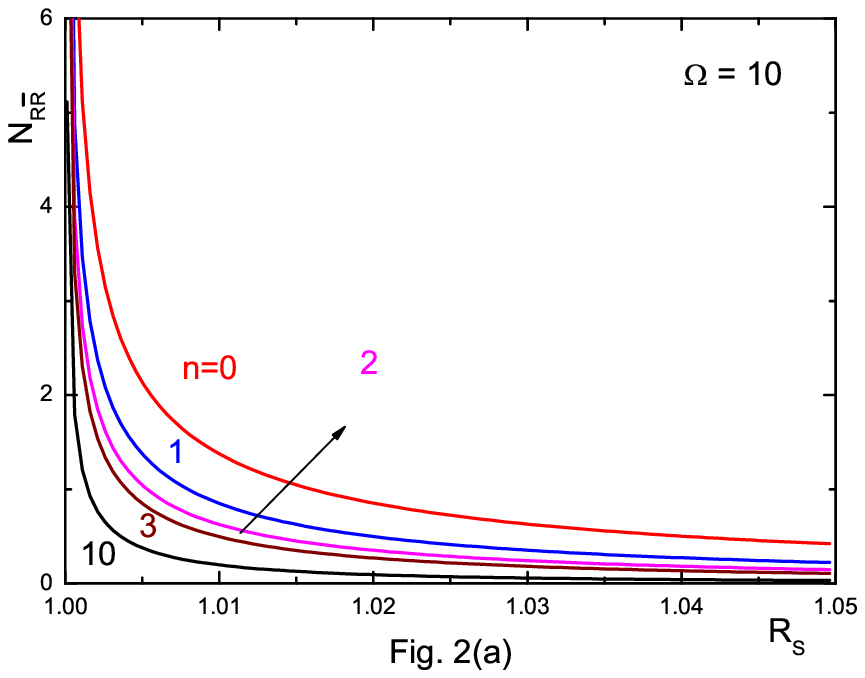}
\includegraphics[height=6.5cm]{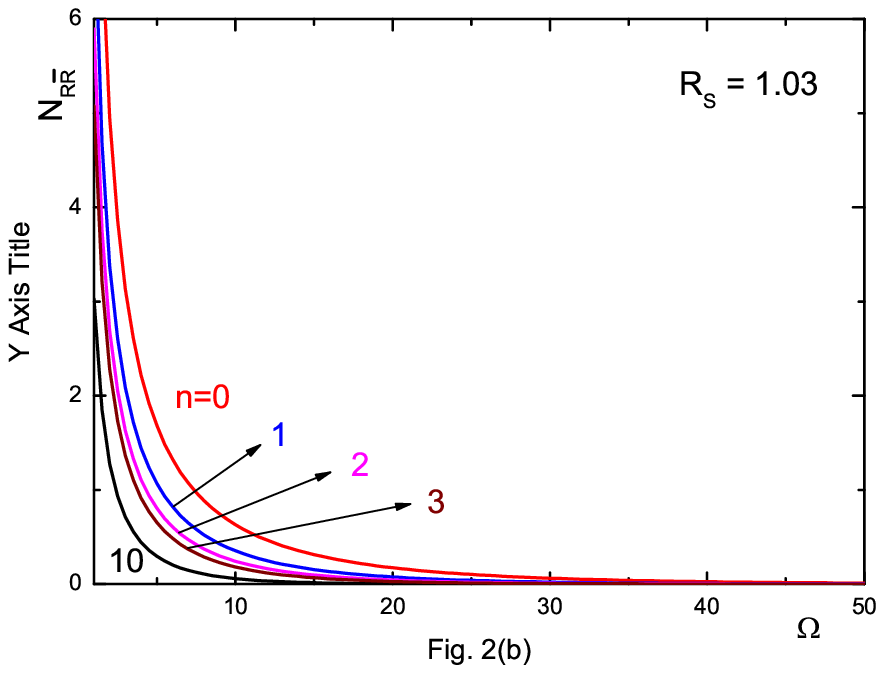}
\caption[fig2]{(Color online) In (a) we plot the $R_S$-dependence of 
${\cal N}_{R\bar{R}}$ when $n=0, 1, 2, 3, 10$. We fix $\Omega$ as 
$\Omega = 10$. In (b) we plot the $\Omega$-dependence of ${\cal 
N}_{R\bar{R}}$ when $n=0, 1, 2, 3, 10$. We fix $R_S$ as 
$R_S = 1.03$. 
}
\end{center}
\end{figure}

In appendix A the negativity ${\cal N}_{R \bar{R}}$ is computed. 
In Fig. 2 we plot ${\cal N}_{R \bar{R}}$ with varying the number of extra dimensions as $n=0, 1, 2, 3, 10$. 
In Fig. 2(a) we plot the $R_S$-dependence of ${\cal N}_{R\bar{R}}$ with choosing $\Omega = 10$. 
In Fig. 2(b) we also plot the $\Omega$--dependence of ${\cal N}_{R\bar{R}}$ with choosing $R_S = 1.03$. 
Contrary to the quantum correlation between Alice and Rob the quantum entanglement between Rob and AntiRob decreases with increasing $n$. 
We do not know why $n$-dependence of ${\cal N}_{R\bar{R}}$ is different from that of ${\cal N}_{AR}$. 
Probably, this is due to the fact that this entanglement is unphysical because Rob is causally disconnected fro AntiRob.
Thus, this quantum correlation is purely theoretical and is useless for the quantum information task.

\section{Conclusion}

In this short paper we compute the various bipartite quantum correlations in the presence of the $(4+n)$-dimensional Schwarzschild 
black hole. In particular, we focus on the $n$-dependence of the bosonic entanglements. For the case between Alice and Rob
the quantum correlation is degraded as expected. The degradation rate decreases with increasing $n$. However, it is 
found that there is no creation of quantum correlation between Alice and AntiRob. For the case between Rob and AntiRob
the quantum entanglement is created although they are separated in the causally disconnected regions. The entanglement is found to 
decrease with decreasing $n$.

In this paper we performed the calculation under the condition of $r_0 \sim r_H$. If we relax this condition, various correlations may
exhibit different behavior. In this case, however, we cannot use the Rindler-anology of the Schwarzschild background. Thus we have to 
re-derive the equations corresponding to Eq. (\ref{inter1}) in this case by computing appropriate Bogoliubov coefficients. 
However, this generalization might be difficult problem mainly due to the difficulty in the analytic computation of the 
Bogoliubov coefficients.

It seems to be of interest to extend this paper to the higher-dimensional rotating black hole case\cite{higher4,higher5}.
The condition for occurring the superradiance in the higher-dimensional rotating black holes was derived in Ref. \cite{super1}. 
It seems to be highly interesting to examine the effect of the superradiance to the entanglement degradation.

{\bf Acknowledgement}:
This work was supported by the Kyungnam University
Foundation Grant, 2012.

\newpage

\begin{appendix}
{\centerline{\bf Appendix A}}
\setcounter{equation}{0}
\renewcommand{\theequation}{A.\arabic{equation}}

In this appendix we compute the various bipartite entanglement.

\subsection{Alice-Rob quantum correlations}

Now, we assume that Rob is in the region where Killing vector is $\partial_t$. Since, then, Rob cannot access the region whose 
time-like Killing vector is $-\partial_t$, one should take a partial trace over $\bar{R}$ in Eq. (\ref{black-1}). Thus, the resulting 
quantum state for Alice-Rob system becomes a mixed state in a form
\begin{eqnarray}
\label{state-AR}
& &\rho_{AR} = \mbox{Tr}_{\bar{R}} \ket{\psi}_{AR\bar{R}}\bra{\psi}                \\   \nonumber
& & = \frac{1}{2} \sum_{n=0}^{\infty} \frac{\tanh^{2n} r}{\cosh^4 r}
\Bigg[ \cosh^2 r \ket{0,n}\bra{0,n} + (n+1) \bigg\{ |q_L|^2 \ket{1,n}\bra{1,n} + |q_R|^2 \ket{1,n+1}\bra{1,n+1} \bigg\}     \\   \nonumber
& & \hspace{3.5cm}
+ \sqrt{n+1} \cosh r \bigg\{ q_R \ket{1,n+1} \bra{0,n} + q_R^* \ket{0,n} \bra{1,n+1} \bigg\}                                 \\   \nonumber
& & \hspace{3.5cm}
+ \sqrt{n+1} \sinh r \bigg\{ q_L \ket{1,n} \bra{0,n+1} + q_L^* \ket{0,n+1} \bra{1,n} \bigg\}                                  \\   \nonumber
& & \hspace{3.0cm}
+ \sqrt{(n+1)(n+2)} \tanh r \bigg\{ q_L q_R^* \ket{1,n} \bra{1,n+2} + q_L^* q_R \ket{1,n+2}\bra{1,n} \bigg\}
\Bigg].
\end{eqnarray}

For $q_R = 1$ computation of ${\cal N}_{AR}$ is straightforward from Eq. (\ref{state-AR}). Since $\rho_{AR}^{T_A}$ can be diagonal 
in terms of $2 \times 2$ matrix if one choose the order of the basis appropriately, the eigenvalues can be easily computed, which is 
$$ \left\{\frac{1}{2 \cosh^2 r}, \Lambda_k^{\pm} \hspace{.3cm} (k = 0, 1, 2, \cdots) \right\}$$     where
\begin{equation}
\label{appen-2}
\Lambda_k^{\pm} = \frac{\tanh^{2k-2} r}{4 \cosh^4 r} \left[ (k + \cosh^2 r \tanh^4 r) \pm 
\sqrt{(k + \cosh^2 r \tanh^4 r)^2 + 4 \cosh^2 r \tanh^4 r} \right].
\end{equation}
Since $\Lambda_k^+ > 0$ and $\Lambda_k^- < 0$, the negativity between Alice and Rob becomes
\begin{equation}
\label{appen-3}
{\cal N}_{AR} = - \sum_{k=0}^{\infty} \Lambda_k^-.
\end{equation}
The extra-dimensional dependence of ${\cal N}_{AR}$ arises from $n$-dependence of $r$ as Eq. (\ref{statistical-1}) shows.

\subsection{Alice-AntiRob quantum correlations}

From Eq. (\ref{black-1}) with $q_R = 1$ the state for Alice-AntiRob becomes
\begin{eqnarray}
\label{state-ARbar}
& &\rho_{A\bar{R}} = \mbox{Tr}_{R} \ket{\psi}_{AR\bar{R}}\bra{\psi}                \\   \nonumber
& & \hspace{1.0cm}
= \frac{1}{2} \sum_{n=0}^{\infty} \frac{\tanh^{2n} r}{\cosh^4 r}
\Bigg[ \cosh^2 r \ket{0,n}\bra{0,n} + (n+1) \ket{1,n}\bra{1,n}                      \\   \nonumber
& & \hspace{3.0cm}     + \sqrt{n+1} \sinh r \bigg\{\ket{0,n+1}\bra{1,n} + \ket{1,n} \bra{0,n+1} \bigg\} \Bigg].
\end{eqnarray}
Since, by similar way, one can show  that all eigenvalues of $\rho_{A\bar{R}}^{T_A}$ are positive, we get ${\cal N}_{A\bar{R}} = 0$. 
Thus, no entanglement is created between Alice and AntiRob.

\subsection{Rob-AntiRob quantum correlations}

From Eq. (\ref{black-1}) with $q_R = 1$ the state for Rob-AntiRob becomes
\begin{eqnarray}
\label{state-RRbar}
& &\rho_{R\bar{R}} = \mbox{Tr}_{A} \ket{\psi}_{AR\bar{R}}\bra{\psi}                \\   \nonumber
& &
= \frac{1}{2} \sum_{m,n=0}^{\infty} \frac{\tanh^{m+n} r}{\cosh^4 r}
\Bigg[\cosh^2 r \ket{n,n}\bra{m,m} + \sqrt{(m+1)(n+1)} \ket{n+1,n}\bra{m+1,m} \bigg].
\end{eqnarray}
Since $\rho_{R\bar{R}}^{T_B}$ is block-diagonal in terms of $m \times m \hspace{.2cm} (m = 1, 2, 3, \cdots)$ matrix provided that the order of the basis is selected as 
$$ \left\{ \ket{00}, \ket{01}, \ket{10}, \ket{02}, \ket{11}, \ket{20}, \cdots \right\},    $$
one can compute the 
eigenvalues in principle. Therefore, it is possible to compute ${\cal N}_{R\bar{R}}$ numerically by making use of Eq. (\ref{negativity1}). 
\end{appendix}


\begin{thebibliography}{99}
\bibitem{bennett93} C. H. Bennett, G. Brassard, C. Cr\'{e}peau, R. Jozsa,
A. Peres and W. K. Wootters, {\it Teleporting an unknown quantum state via dual classical and Einstein-Podolsky-Rosen channels},  
Phys. Rev. Lett. {\bf 70} (1993) 1895.
\bibitem{cryptography} C. H. Bennett and G. Brassard, {\it Quantum Cryptography: Public Key Distribution and Coin Testing}, 
in Proceedings of the IEEE International Conference on
Computer, Systems, and Signal Processing, Bangalore, India (IEEE, New York, 1984), pp. 175-179;
A. K. Ekert, {\it Quantum cryptography based on Bell’s theorem}, Phys. Rev. Lett. {\bf 67} (1991) 661;
 C. A. Fuchs, N. Gisin, R. B. Griffiths, C. S. Niu and A. Peres, {\it Optimal eavesdropping in quantum cryptography. I. 
Information bound and optimal strategy}, Phys. Rev. {\bf A 56} (1997) 1163 [quant-ph/9701039].
\bibitem{qc} G. Vidal, Phys. Rev. Lett. {\it Efficient Classical Simulation of Slightly Entangled Quantum Computations}, 
{\bf 91} (2003) 147902 [quant-ph/0301063];
M. A. Nielsen and I. L. Chuang, {\it Quantum Computation and
Quantum Information} (Cambridge University Press, Cambridge, England, 2000).
\bibitem{horodecki07} R. Horodecki, P. Horodecki, M. Horodecki, and K. Horodecki, {\it Quantum entanglement}, 
Rev. Mod. Phys. {\bf 81} (2009) 865 [quant-ph/0702225] and references therein.
\bibitem{epr-35} A. Einstein, B. Podolsky and N. Rosen, {\it Can quantum-mechanical description of physical 
reality be considered complete ?}, Phys. Rev. {\bf A47} (1935) 777.
\bibitem{schrodinger-35} E. Schr\"{o}dinger, {\it Die gegenw\"{a}rtige Situation in der Quantenmechanik}, Naturwissenschaften, 
{\bf 23} (1935) 807.
\bibitem{peres1} 
A. Peres and D. R. Terno, {\it Quantum information and relativity theory}, Rev. Mod. Phys. {\bf 76} (2004) 93 [quant-ph/0212023].
\bibitem{peres2}
A. Peres and D. R. Terno, {\it Quantum Information and Special Relativity}, Int. J. Quant. Info. {\bf 1} (2003) 225 [quant-ph/0301065].
\bibitem{peres3}
A. Peres, {\it Quantum information and general relativity}, Fortsch. Phys. {\bf 52} (2004) 1052 [quant-ph/0405127]. 
\bibitem{inertial1} 
M. Czachor, {\it Einstein-Podolsky-Rosen-Bohm experiment with relativistic massive particles}, 
Phys. Rev. {\bf A 55} (1997) 72 [quant-ph/9609022].
\bibitem{inertial2} 
A. Peres, P. F. Scudo and R. Terno, {\it Quantum Entropy and Special Relativity}, Phys. Rev. Lett. {\bf 88} (2002) 230402 [quant-ph/0203033].
\bibitem{inertial3}
P. M. Alsing and G. J. Milburn, {\it Lorentz invariance of entanglement}, Quantum Inf. Comput. {\bf 2} (2002) 487 [quant-ph/0203051].
\bibitem{inertial4} 
R. M. Gingrich and C. Adami, {\it Quantum Entanglement of Moving Bodies}, Phys. Rev. Lett. {\bf 89} (2002) 270402 [quant-ph/0205179].
\bibitem{noninertial1} 
I. Fuentes-Schuller and R. B. Mann, {\it Alice Falls into a Black Hole: Entanglement in Noninertial Frames}, 
Phys. Rev. Lett. {\bf 95} (2005) 120404 [quant-ph/0410172].
\bibitem{noninertial2}
Y. Ling, S. He, W. Qiu and H. Zhang, {\it Quantum entanglement of electromagnetic field in non-inertial reference frames}, 
J. Phys: Math. Theor. {\bf A40} (2007) 9025 [quant-ph/0608029].
\bibitem{noninertial3}
Q. Pan and J. Jing, {\it Degradation of nonmaximal entanglement of scalar and Dirac fields in noninertial frames}, 
Phys. Rev. {\bf A 77} (2008) 024302 [arXiv:08021238 (quant-ph)].
\bibitem{noninertial4}
A. Datta, {\it Quantum discord between relatively accelerated obserbers}, Phys. Rev. {\bf A 80} (2009) 052304 [arXiv:0905.3301 (quant-ph)].
\bibitem{noninertial5}
J. Wang, J. Deng, and J. Jing, {\it Classical correlation and quantum discord sharing of Dirac fields in noninertial frames}, 
Phys. Rev. {\bf A 81} (2010) 052120 [arXiv:0912.4129 (quant-ph)].
\bibitem{noninertial6}
M. R. Hwang, D. K. Park, and E. Jung, {\it Tripartite entanglement in a noninertial frame}, 
Phys. Rev. {\bf A 83} (2011) 012111 [arXiv:1010.6154 (hep-th)].
\bibitem{noninertial7}
E. Martin-Martinez and I. Fuentes, {\it Redistribution of particle and antiparticle entanglement in noninertial frames}, 
Phys. Rev. {\bf A 83} (2011) 052306 [arXiv:1102.4759 (quant-ph)].
\bibitem{noninertial8}
M. Montero and E. Martin-Martinez, {\it Fermionic entanglement ambiguity in noninertial frames}, 
Phys. Rev. {\bf A 83} (2011) 062323 [arXiv:1104.2307 (quant-ph)].
\bibitem{noninertial9}
M. Montero, J. Leon, and E. Martin-Martinez, {\it Fermionic entanglement extinction in noninertial frames}, 
Phys. Rev. {\bf A 84} (2011) 042320 [arXiv:1108.1111 (quant-ph)].
\bibitem{noninertial10}
A. Smith and R. B. Mann, {\it Persistence of Tripartite Nonlocality for Non-inertial observers}, arXiv:1107.4633 (quant-ph).
\bibitem{noninertial11}
D. K. Park, {\it Tripartite Entanglement-Dependence of Tripartite Non-locality in Non-inertial Frame},  arXiv:1201.1335 (quant-ph).
\bibitem{noninertial12}
Mi-Ra Hwang, Eylee Jung, and DaeKil Park, {\it Three-Tangle in Non-inertial Frame}, will appear in Class. Quantum Grav.
[arXiv:1203.5594 (quant-ph)].
\bibitem{unruh1} 
W. G. Unruh, {\it Notes on black-hole evaporation}, Phys. Rev. {\bf D 14} (1976) 870.
\bibitem{unruh2}
 N. D. Birrel and P. C. W. Davies, {\it Quantum Fields in Curved Space} (Cambridge University Press, Cambridge, England, 1982).
\bibitem{black1} Q. Pan and J. Jing, {\it Hawking radiation, entanglement, and teleportation in the background of an asymptotically 
flat static black hole}, Phys. Rev. {\bf D 78} (2008) 065015 [arXiv:0809.0811 (gr-qc)].
\bibitem{black2} 
E. Mart\'in-Mart\'inez, L. J. Garay, and J. Le\'on, {\it Unveiling quantum entanglement degradation near a Schwarzschild black hole}, 
Phys. Rev. {\bf D 82} (2010) 064006 [arXiv:1006.1394 (quant-ph)].
\bibitem{black3} Eylee Jung, Mi-Ra Hwang, and DaeKil Park, {\it Quantum Discord and Quantum Entanglement in the Background of an Asymptotically Flat Static Black Holes},  arXiv:1206.5608 (hep-th).
\bibitem{brane1} N. Arkani-Hamed, S. Dimopoulos and G. Dvali, {\it The Hierarchy Problem and New Dimensions at a Millimeter}, 
Phys. Lett. {\bf B 429} (1998) 263 [hep-ph/9803315].
\bibitem{brane2} L. Antoniadis, N. Arkani-Hamed, S. Dimopoulos and G. Dvali, {\it New Dimensions at a Millimeter to a Fermi and Superstrings at a TeV}, Phys. Lett. {\bf B 436} (1998) 257 [hep-ph/9804398].
\bibitem{brane3} L. Randall and R. Sundrum, {\it A Large Mass Hierarchy from a Small Extra Dimension}, Phys. Rev. Lett. 83 (1999) 3370 [hep-ph/9905221].
\bibitem{brane4} L. Randall and R. Sundrum, {\it An Alternative to Compactification}, Phys. Rev. Lett. 83 (1999) 4690 [hep-th/9906064].
\bibitem{higher1} C. M. Harris and P. Kanti, {\it Hawking Radiation from a $(4+n)$-dimensional Black Hole: Exact Results for the Schwarzschild Phase}, JHEP {\bf 0310} (2003) 014 [hep-ph/0309054].
\bibitem{higher2} E. Jung and D. K. Park, {\it Absorption and Emission Spectra of an higher-dimensional Reissner-N\"ordström black hole}, Nucl. Phys. {\bf B 717} (2005) 272 [hep-th/0502002].
\bibitem{higher3} E. Jung and D. K. Park, {\it Bulk versus Brane Emissivities of Photon Fields: For the case of Higher-Dimensional Schwarzschild Phase}, Nucl. Phys. {bf B 766} (2007) 269 [hep-th/0610089].
\bibitem{higher4} V. Frolov and D. Stojkovi´c, {\it Quantum radiation from a $5$-dimensional rotating black hole}, 
Phys. Rev. {\bf D 67} (2003) 084004 [gr-qc/0211055].
\bibitem{higher5} E. Jung and D. K. Park, {\it Bulk versus Brane in the Absorption and Emission : $5D$ Rotating Black Hole Case}, 
Nucl. Phys. {\bf B 731} (2005) 171 [hep-th/0506204].
\bibitem{negativity} G. Vidal and R.F. Werner, {\it A computable measure of entanglement}, Phys. Rev. {\bf A 65} (2002) 032314 [quant-ph/0102117].
\bibitem{myers86} R. C. Myers and M. J. Perry, {\it Black Holes in Higher
Dimensional Space-Times}, Ann. Phys. {\bf 172} (1986) 304.
\bibitem{boulware75} D. G. Boulware, {\it Quantum field theory in Schwarzschild and Rindler spaces}, Phys. Rev. {\bf D 11} (1975) 1404.
\bibitem{beyond} D. E. Bruschi {\it et al}, {\it Unruh effect in quantum information beyond the single-mode approximation}, Phys. Rev. {\bf A 82} (2010) 042332 [arXiv:1007.4670 (quant-ph)].
\bibitem{bennett96} C. H. Bennett, D. P. DiVincenzo, J. A. Smolin and W. K. Wootters,
{\it Mixed-state entanglement and quantum error correction}, Phys. Rev. {\bf A54}
(1996) 3824 [quant-ph/9604024].
\bibitem{woot-98}W. K. Wootters, {\it Entanglement of Formation of an Arbitrary State of
Two Qubits}, Phys. Rev. Lett. {\bf 80}, 2245 (1998) [quant-ph/9709029].
\bibitem{vedral-97-1} V. Vedral, M. B. Plenio, M. A. Rippin and P. L. Knight, {\it Quantifying 
Entanglement}, Phys. Rev. Lett. {\bf 78} (1997) 2275 [quant-ph/9702027].
\bibitem{super1} Eylee Jung, SungHoon Kim, D. K. Park, {\it Condition for Superradiance in Higher-dimensional Rotating Black Holes}, 
Phys. Lett. {\bf B 615} (2005) 273 [hep-th/0503163]; {\it Condition for the Superradiance Modes in Higher-Dimensional 
Rotating Black Holes with Multiple Angular Momentum Parameters}, {\it ibid}, {\bf B 619} (2005) 347 [hep-th/0504139].
\end{thebibliography}
\end{document}